\documentclass[aps,pra,twocolumn,showpacs,superscriptaddress,groupedaddress]{revtex4}  % for review and submission
\usepackage{graphicx}  % needed for figures
\usepackage{dcolumn}   % needed for some tables
\usepackage{bm}        % for math
\usepackage{amssymb}   % for math

\usepackage{amsmath}

\usepackage{hyperref}

\begin{document}

% The following information is for internal review, please remove them for submission
\widetext
%\leftline{Version xx as of \today}
%\leftline{Primary authors: Joe E. Physics}
%\leftline{To be submitted to (PRL, PRD-RC, PRD, PLB; choose one.)}
%\leftline{Comment to {\tt d0-run2eb-nnn@fnal.gov} by xxx, yyy}
%\centerline{\em D\O\ INTERNAL DOCUMENT -- NOT FOR PUBLIC DISTRIBUTION}

% the following line is for submission, including submission to the arXiv!!
%\hspace{5.2in} \mbox{Fermilab-Pub-04/xxx-E}

\title{Electron dynamics in the carbon atom induced by spin-orbit interaction}
\author{H.F. Rey}
\email{h.rey@qub.ac.uk}
\author{H.W. van der Hart}

%\homepage{http://stoa.usp.br/thschiavo}
\affiliation{Centre for Theoretical Atomic, Molecular and Optical Physics, School of
Mathematics and Physics\\ Queen's University Belfast, Belfast BT7 1NN, United Kingdom}

%\input author_list.tex       % D0 authors (remove the first 3 lines
                             % of this file prior to submission, they
                             % contain a time stamp for the authorlist)
                             % (includes institutions and visitors)
\date{\today}

\begin{abstract}
We use R-Matrix theory with Time dependence (RMT) to investigate multiphoton ionization of
ground-state atomic carbon with initial
orbital magnetic quantum number $M_L$ = 0 and $M_L$ = 1 at a laser wavelength of 390 nm and
peak intensity of 10$^{14}$ W cm$^{-2}$. Significant differences in
ionization yield and ejected-electron momentum distribution are observed between the two
values for $M_L$. We use our theoretical
results to model how the spin-orbit interaction affects electron emission along the laser
polarization axis. Under the assumption that
an initial C atom is prepared at zero time delay with $M_L$=0, the dynamics with
respect to time delay of an ionizing probe pulse modeled
using RMT theory is found to be in good agreement with available experimental data. 
\end{abstract}

\maketitle

%\section{\label{sec:level1}Introduction}
% sections are not used for PRL papers

\section{Introduction}

Over the last 15 years great advances have been made in the development and application
of experimental laser techniques on the sub-femtosecond time-scale
(1 fs = 10$^{-15}$ s) \cite{Kra09}. These advances have the prospect of the detailed
study and potential guiding of electron motion in atoms and molecules.
Several methods have been applied to monitor the electron dynamics, such as pump-probe
spectroscopy using few-cycle pulses, through which, for
example, the angular correlation between sequential ionization steps was investigated
\cite{Fle11}. Another example is transient absorption spectroscopy, in
which, for example, sequential ionization dynamics of Kr was investigated \cite{Gou10}.
This experiment demonstrated the influence of the spin-orbit interaction
on the Kr$^+$ dynamics when sufficiently short ionization pulses are employed. 

Electron motion in a general atom can be regarded as the complement of atomic structure
via the time-energy uncertainty principle. In the atomic structure of light
atoms, relativistic interactions, such as spin-orbit coupling, are normally assumed to be
negligible. However, in heavier atoms, this assumption no longer holds. As a
consequence, it can be assumed that spin-orbit dynamics may have a relatively minor effect
on electron dynamics of light atoms on the time scale of a
short laser pulse, whereas this dynamics will become  important when heavier atoms are
considered. That does not mean, however, that spin-orbit dynamics is
irrelevant for light atoms: when light atoms interact with a sequence of two light pulses
with a long time delay between the two, spin orbit dynamics can affect the intermediate
state significantly \cite{Hul13,Ekl13}.

Changes in the atomic state induced by spin-orbit interaction can affect the subsequent
atomic dynamics significantly \cite{Sok00}. Total angular momentum
$J=L+S$, with $L$ total orbital angular momentum and $S$ total spin, is conserved, as
is its projection on the $z$-axis, $M_J= M_L + M_S$. However,
although $M_J$ is conserved, its components $M_L$ and $M_S$ are not. Hence, the spin-orbit
interaction can change, for example, a $JM_JLS$ state with $M_L=0$
and $M_S=+1/2$ into a $JM_JLS$ state with $M_L=+1$ and $M_S=-1/2$. This change in orbital
magnetic quantum number can have a noticeable effect on
the atomic dynamics: For example, in harmonic generation of noble-gas ions Ne$^+$ and
Ar$^+$, the harmonic yields for $M_L=0$ were reduced by a factor 2-4 compared to $M_L=\pm1$
\cite{Bro13,Has14}.  

Recently, pump-probe experiments have been carried out to investigate the spin-orbit induced
dynamics in the C ground state following photodetachment of C$^-$~\cite{Hul13,Ekl13}. In these
studies, an initial C atom was formed by photodetachment of C$^-$. This residual C atom is
left in a superposition of ground state $^3$P$^e$ levels with different $J$. The energy
splitting between the different $J$ levels then leads to dynamics within the different
$M_L$ levels, which can be observed through differences in the measured ejected-electron
momentum spectra between parallel polarization of the pump and probe pulses and perpendicular
polarization of the pulses. These differences were shown to vary with a period of 760 to 2000
fs, as can be deduced from the
energy difference between the $J$ levels of the C $^3$P$^e$ ground state, 2.0 meV between the
lower $J=0$ and the $J=1$ level, and 3.4 meV between the $J=1$ and $J=2$ level \cite{Moo93}. 
 
From a theoretical perspective, the influence of relativistic interactions on ultra-fast atomic
dynamics has already been the subject of theoretical investigation \cite{Roh09,Pab12}. However,
it would
be very useful to develop this capability in other theoretical methods and techniques as well,
such as, for example, time-dependent R-matrix theory \cite{Lam08,Lys09,Lys11,Moo11}.  The most
recent implementation of time-dependent R-matrix theory, named R-matrix theory including time
dependence (RMT), is the most efficient implementation for large-scale studies, as it provides
better stability when many angular momenta are included and provides better scope for exploitation
of massively parallel computing facilities. The first step in this development is to verify that
initial states with $M_L\neq0$ can be investigated accurately using the RMT approach. However,
whereas the previous implementation of time-dependent R-matrix theory has already been applied
to the investigation of dynamics for atoms with $M_L\neq0$ \cite{Hut11,Bro13,Has14}, up to now
RMT theory has only been applied to systems with $M_L=0$ \cite{Moo11b,Har14}.

In this report, we demonstrate that RMT theory can be applied to the study of atoms with an
initial $M_L\neq0$. We apply the theory to investigate ejected-electron momentum distributions
for the multiphoton ionization of C atoms in short 390-nm laser pulses with a peak intensity of
10$^{14}$ W/cm$^2$ for both $M_L=0$ and $M_L=1$. We will explain the significant changes in these
momentum distributions between the two values of $M_L$. We then use these momentum distributions
to model the observation of spin-orbit dynamics in C as a function of time delay $\tau$ for the
ionization pulse, under the assumption that a ground-state C atom is created at time $t=0$ by
emission of an $m_\ell=0$ electron from ground-state C$^-$. Since the spin-orbit dynamics in C
occurs on timescales longer than the duration of the 390-nm laser pulse, which takes 1.3 fs per
cycle, it can be assumed that
no spin-orbit dynamics occurs during the interaction with the laser pulse. The time-delayed
initial state of C can then be regarded as a superposition of states with $M_L=1,$ $M_L=0$,
and $M_L=-1$.

The organisation of the report is as follows. In Sec. \ref{sec:theory}, we give a brief overview
of RMT theory, including a description of changes required to treat non-zero $M$. The specific
parameters used in the present calcutations, including a description of the C$^{+}$ basis set,
are also included. In Sec. \ref{sec:results}, we present our calculated ejected-electron momentum
distributions along with seleceted photoelectron energy spectra focusing on the differences seen
for $M_L=0$ and $M_L=1$. We then model  the differences in the ejected-electron momentum spectra
that would be induced by spin orbit dynamics, and compare these with the experimental findings
\cite{Hul13,Ekl13}. Finally in Sec \ref{sec:conclusions}, we report our conclusions.

%\subpart{Template information}

\section{Methods}
\label{sec:theory}

\subsection{The RMT approach}

In this report, we employ the RMT approach to investigate multiphoton ionization of neutral carbon
and model the ejected electron dynamics.
The RMT method builds upon traditional R-matrix theory~\cite{Bur93,Bur11}, by separating the
description of the system into two distinct regions: an inner
region and an outer region. In the inner region, the wavefunction is expressed in terms of 
a multi-electron R-matrix basis description within
a confined spatial region surrounding the nucleus \cite{Bur11}. In the outer region, the full
multi-electron wavefunction is descibed as a direct product of a
channel function and a finite-difference representation of the single-electron radial wavefunction
for an outer elecron found only outside the confined
spatial region \cite{Lys11,Moo11}. The channel function comprises the wavefunction of a residual-ion
state coupled with the spin and angular momentum functions of the
ejected electron.

A key feature of the RMT approach is the connection between the inner and outer regions. To provide
information about the inner-region wavefunction to the outer
region, the outer-region grid is extended into the inner region. The inner-region radial wavefunction
is then evaluated on this grid extension, and made available to 
the outer region.  In order to provide outer-region wavefunction information to
the inner-region, time derivatives of the outer-region wavefunction are evaluated at the inner-region
boundary, and are then provided to the inner-region. Full details about
the link between the two regions are provided in \cite{Lys11,Moo11}. 

Time propagation within the RMT approach is carried out through Arnoldi propagators \cite{Smy98}.
Through the grid extension into the inner region, all the necessary
information for propagation of the outer-region wavefunction is available on the outer-region grid,
and the propagation through an Arnoldi propagator is relatively
straightforward. In the inner region, however, different Arnoldi propagators are needed for the
wavefunction, and for each outer-region time derivative at the
boundary. All time derivatives are therefore propagated separately, and the full time-propagated
wavefunction is obtained by combining all the separate propagation
terms \cite{Lys11,Moo11}.

The implementation of the RMT codes to date was limited to systems with an initial angular momentum
$L=0$ and, consequently, $M_L = 0$. However, the ground state of
C has $^3$P$^e$ symmetry which can have a total magnetic quantum number $M_L=-1,0,$ or $1$. The
value of $M_L$ significantly affects the calculations. According to dipole
selection rules, only transitions with $\Delta L$~= $\pm 1$ are allowed for $M_L$ = 0. On the other
hand, for $M_L$ = 1, transitions with $\Delta L$~=~$0$ are allowed in addition
to $\Delta L$ = $\pm 1$. As a consequence, for $M_L=0$, only a single parity needs to be included
for each value of $L$, whereas for $M_L$ = 1, both parities need to be included
for each value of $L$. More symmetries are therefore available for $M_L=1$ and the calculations
approximately double in size. In addition, we need to take into account that all symmetries can
now potentially interact with three other symmetries, rather than two. Within the parallel
implementation of the RMT codes, this requires additional communication between MPI tasks. However,
since the propagation involves only matrix-vector multiplications, no reordering of the symmetry
blocks is required, in contrast to the previous implementation of time-dependent R-matrix theory
\cite{Hut11}. 

\subsection{The description of C}

To study the influence of the initial value of $M_L$ values on the multiphoton ionization dynamics of
C, we employ an R-matrix basis set previously employed to investigate
photoionization of C \cite{Bur79}. In this basis set, all possible residual ion states are considered
that can be formed using 2s and 2p orbitals:
\begin{eqnarray*}
			n\hbar\omega + {\rm C} (2s^{2}2p^{2}\ {}^{3}\textrm{P}^{e})\! & \! 
			\rightarrow \! & \! {\rm C}^{+} (2s^{2}2p\   { }^{2}\textrm{P}^{o}) + e^{-} \\
&\! \rightarrow \!& \!{\rm C}^{+} (2s2p^{2}\  { }^{2,4}\textrm{P}^{e}, { }^{2}\textrm{D}^{e}, 
{ }^{2}\textrm{S}^{e}) + e^{-} \\
&\! \rightarrow \!& \!{\rm C}^{+} (2p^{3}\   { }^{4}\textrm{S}^{o}, { }^{2}\textrm{D}^{o},
{ }^{2}\textrm{P}^{o}) + e^{-}
\end{eqnarray*}
The 1s, 2s and 2p orbitals building these three ionic configurations are given by the 1s, 2s and 2p
Hartree-Fock orbitals for the $^{2}P^{o}$ ground state of the ion \cite{CR74}.
To improve the description of the C$^+$ eigenstates, the basis set also includes $\overline{3s}$, 
$\overline{3p}$, and $\overline{3d}$ pseudo-orbitals orthogonal to the Hartree-Fock orbitals. The
C$^+$ basis set is then expanded by allowing single excitations from the 2s$^2$2p, 2s2p$^2$ and
2p$^3$ configurations to these pseudo-orbitals. Energies of the lowest five eigenstates of C$^+$
are given in table \ref{tab:cplusdata} and compared with literature values \cite{Moo93}. The
energies of the 2p$^3$ states are not given, as they lie above the 2s$^2$3s states, which have
been excluded from the present calculations. The energies agree exactly with those obtained in
\cite{Bur79}, apart from the energy for the 2s2p$^2$ $^2$S$^e$ state. This is most likely due to
a small difference in the CI expansion used for this
state, which includes 6 basis functions in the present calculations, whereas the one reported
in \cite{Bur79} includes 5 basis functions. The table demonstrates that the lowest two thresholds
are within 0.2 eV of the literature values, whereas the differences increase rapidly for the higher
states. This is not surprising as the 2s2p$^2$ $^2$D$^e$ and $^2$S$^e$ states will interact strongly
with the physical 2s$^2$3d and 2s$^2$3s states, respectively. The $\overline{3s}$ and $\overline{3d}$
orbitals used in the present calculations are pseudo-orbitals, and will therefore not accurately reflect
the 3s and 3d orbitals in the physical states. Hence the interaction with 2s$^2$3d and 2s$^2$3s will not be
described accurately, and the obtained threshold energies will be
less accurate. Nevertheless, the most important threshold states in the present multiphoton calculations
are the lowest two thresholds, and these are
described with sufficient accuracy for the present purposes.

\begin{table}
\caption{ C$^+$ threshold energies relative to the ground state of C, compared
to literature values \cite{Moo93}. Energies are not given for the 2p$^3$ states
as these lie inbetween the 2s$^2$3$\ell$ and
2s$^2$4$\ell$ states. Apart from the energy for the 2s2p$^2$ $^2$S$^e$ state, all present
energies agree with those in \cite{Bur79}.} 
\begin{tabular}{l|cc}
C$^+$ state & Present energy & Literature \cite{Moo93} \\
 & eV & eV \\ \hline 
2s$^2$2p $^2$P$^o$ & 11.366 & 11.262 \\
2s2p$^2$ $^4$P$^e$ & 16.430 & 16.592 \\
2s2p$^2$ $^2$D$^e$ & 20.962 & 20.547 \\
2s2p$^2$ $^2$S$^e$ & 24.196 & 23.220 \\
2s2p$^2$ $^2$P$^e$ & 25.706 & 24.976\\ 
\end{tabular}
\label{tab:cplusdata}
\end{table}

The C basis set is formed by combining the eigenstates for C$^+$ obtained given the basis set
above with a basis set describing the continuum electron. The functions in this  basis set are
described in terms of 65 B-splines of order $k=11$. This basis set is augmented by correlation
functions, which are created by adding one of the input orbitals to the configurations used in
the description of C$^+$. It is further ensured that every function in the continuum basis set
is orthogonal to any input orbital.  The radius of the inner region was chosen to be 27 $a_0$.
This is larger than considered in the previous C photoionization calculations, as, at present,
the initial state needs
to be fully contained within the inner region in the RMT approach. The C ground-state energy
has not been shifted.

In the multiphoton ionization calculations, we investigate C atoms irradiated by a laser field with
a wavelength of 390 nm. The total pulse length is 8 cycles, consisting of a 3-cycle
sin$^2$ ramp-on, 2 cycles at peak intensity, followed by a 3-cycle sin$^2$ ramp-off of the electric
field. Following the end of the pulse, the wavefunction is propagated field-free for another twelve
cycles. For the present investigations, the peak intensity was chosen to be 10$^{14}$ W/cm$^2$. 

In the analysis of the final wavefunction, we only consider the outer region wavefunction, as continuum
electrons are assumed to have escaped the inner region after the propagation at the end of the pulse.
Within the RMT approach, the outer region wavefunction is decribed in terms of a residual ionic state
plus spin and angular momentum of the outer electron coupled with a radial wavefunction for the outer
electron on a discretized finite difference grid. The finite-difference grid has a grid spacing of
$\Delta r=0.15$ $a_0$, and extends to an outer radius of over 3900 $a_0$. 

As indicated earlier, time propagation within the RMT approach is achieved through the use of Arnoldi
propagators \cite{Smy98}. In the present application, we use Arnoldi propagators of
order 12 with a time step of  $\Delta t$ = 0.01 atomic units.

\section{Results}
\label{sec:results}

The main goal in the present calculations is to study how spin-orbit dynamics affects the multiphoton
ionization characteristics of ground-state carbon atoms.
Since the spin-orbit dynamics in ground-state C
is slow compared to the present photoionization dynamics, we can decouple the
two types of dynamics and instead investigate how changes in the
orbital magnetic quantum number $M_L$ are reflected in the multiphoton ionization characteristics.
We can subsequently investigate how spin-orbit dynamics changes the magnetic sublevel populations in
time, and then derive the changes in the ejected-electron momentum spectra driven by the spin-orbit
interaction.

A second objective of the present calculations is to investigate the convergence of the final-state
populations in
the different partial waves with the number of angular momenta retained in the calculation. We adopted
as our convergence criterion that the final population in each outer
channel should have converged to within 0.01\% of the population in the most populated channel. This
convergence was achieved for a maximum angular momentum to be retained in the calculation,
\textit{L}\(_{\rm max}\), of 53. We also carried out calculation with different propagation orders,
and found no significant difference between propagation orders
of 12 and 14. All calculations were therefore carried using an Arnoldi propagator of order 12. 

\begin{figure}%[h!]
	\centering
		\includegraphics[width=9cm]{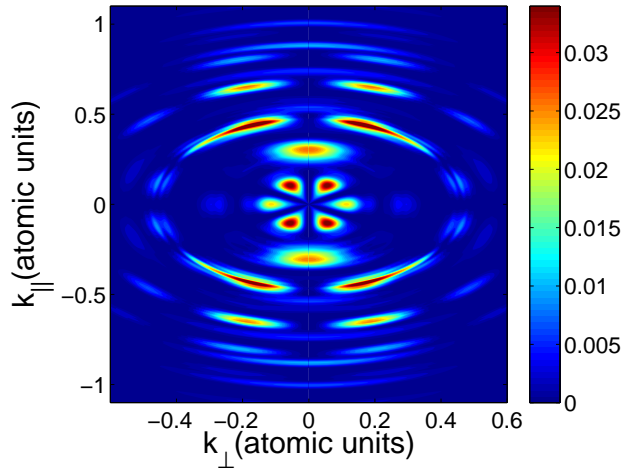}%[width=150pt]
	\caption{(Color online) Ejected-electron momentum distribution in the $k_xk_z$-plane for C
	initially in the ground state with $M_L$ = 0, irradiated by an 8 cycle UV laser pulse with
	wavelength 390 nm and intensity 10$^{14}$ W/cm$^2$. The distribution includes
	the emission of electrons towards the lowest two states of C$^+$,
	which contribute over 99\% to the total
	outer-region population in the present calculations.}
\label{fig:elecmom0}
 \end{figure}

\begin{figure}%[h!]
	\centering
		\includegraphics[width=9.0cm]{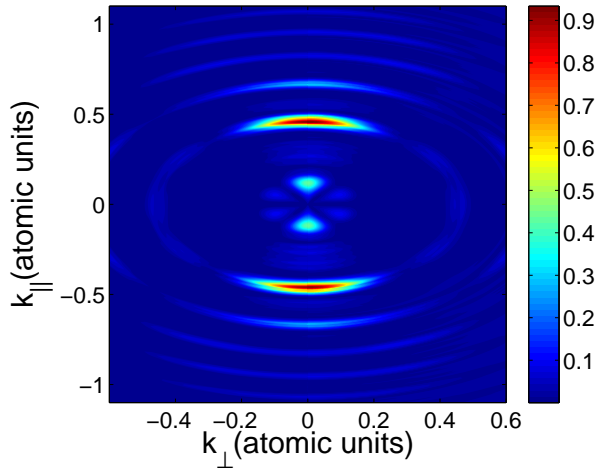}%[width=150pt]
	\caption{(Color online) Ejected-electron momentum distribution in the $k_xk_z$-plane for C
	initially in the ground state with $M_L$ = 1, irradiated by an 8 cycle UV laser pulse with
	wavelength 390 nm and intensity 10$^{14}$ W/cm$^2$. The distribution includes
	the emission of electrons towards the lowest two states of C$^+$,
	which contribute over 99\% to the total
	outer-region population in the present calculations.}
\label{fig:elecmom1}
 \end{figure}

The main outcomes of the calculations are shown in figures \ref{fig:elecmom0} and \ref{fig:elecmom1},
which show the ejected-electron momentum distributions obtained for ground-state
C atoms irradiated by 390 nm light at an intensity of 10$^{14}$ W/cm$^2$ for $M_L=0$ and
$M_L=1$, respectively. The determination of
the ejected-electron momentum distributions is, however, not straightforward
at this intensity. At an approximate intensity of 0.95$\times10^{13}$ W/cm$^2$, channel closing
occurs, so that five photons are needed for ionization instead of four. However, if the
bandwidth of the laser pulse is taken into account as well, both four-photon excitation
of Rydberg states and four-photon ionization will occur simultaneously. To improve the separation
of Rydberg-state population from continuum-state population, the system is propagated for
twelve 390-nm cycles after the pulse has ended.

In the experimental analysis of the ejected-electron momentum distributions, however, the
low-energy part of the distributions was not taken into account \cite{Hul13,Ekl13}. We have therefore
chosen to do the same in the present analysis. At the end of the calculations, we obtain
the outer-region part of the final-state wavefunction. The outer electron is decoupled
from the residual ion, which enables us to extract a wavefunction for the outer electron
associated with each residual-ion state. We then transform the wavefunction for the
outer electron for distances larger than 108 $a_0$ into the momentum representation under
the assumption that the Coulomb potential can be neglected. This distance was chosen by
examination of the final-state wavefunction. For distances greater than 108 $a_0$, the final-state
wavefunction has clear continuum-wave characteristics, whereas this is not the case for
distances smaller than 108 $a_0$. To examine the influence of our choice of distance, we have
carried out our analysis for other distances as well. We observe
that, if C$^+$ is left in the ground
state, this leads to quantitative convergence of the momentum spectra for momenta greater
than about 0.30 atomic units, with a good qualitative description for lower momenta. The momentum
of an electron absorbing five photons is about 0.48 atomic units, and this part of the spectrum of
primary interest is not affected by our choice of distance. If C$^+$ is
left in the lowest excited state, the procedure leads to a good quantitative description
over the entire momentum spectrum due to the absence of threshold photoelectrons. The
momentum of an electron absorbing the minimum number (six) of photons is about 0.30 atomic units.

Since the C ground state has $^3$P$^e$ symmetry, figures \ref{fig:elecmom0} and \ref{fig:elecmom1}
show the momentum
distributions for an initial orbital magnetic quantum number $M_L=0$ and
$M_L=1$, respectively. The figure clearly shows great differences between the two momentum distributions
associated with the change of magnetic quantum number. For $M_L=1$, almost all the ionization
is aligned along the laser polarization axis, whereas for $M_L=0$, two different contributions
can be seen, one aligned along the laser polarisation axis and one at an angle to it. 

Two different distributions can be seen in the ejected-electron momentum distributions for an
initial $M_L=0$ in figure \ref{fig:elecmom0}. These are associated with different residual-ion states of C$^+$, or, alternatively,
the ejection of different electrons. Emission along the laser polarization axis involves
emission of a 2s electron, leaving C$^+$ in the excited 2s2p$^2$ $^4$P$^e$ state, whereas
emission away from the laser polarization axis involves emission of a 2p electron leaving C$^+$
in its 2s$^2$2p $^2$P$^o$ ground state. For $M_L=0$, the total population beyond a distance of 108 $a_0$, 
used in the determination of figure \ref{fig:elecmom0}, in 2p emission channels is 2.86\%, whereas the
one in channels leaving C$^+$ in the 2s2p$^2$ $^4$P$^e$ state
is 0.55\%. The total population in the outer region at the end of the calculation is 4.73\%, with
4.16\% associated with the C$^+$ ground state and 0.56\% with the first excited state of C$^+$.
Higher-lying states of C$^+$ account for less than 0.4\% of the total population in the outer
region, and therefore the contribution to the momentum distributions from outgoing electrons
attached to these states is assumed to be negligible. These populations indicate that although
the majority
of the population in the outer region is assigned to ionization,
a significant amount of population may be associated with Rydberg-state population. The main reason
for this is the proximity of the intensity used in the calculations to a channel-closing
intensity.

The same two distributions also make up the ejected-electron momentum distributions for an initial
magnetic quantum number $M_L=1$, shown in figure \ref{fig:elecmom1}.
However, in this case, the momentum distribution is dominated by
the emission of a 2p electron along the laser polarization axis. Emission of a 2s electron towards
the 2s2p$^2$ state of C$^+$ is still possible, but it is not apparent in the present ejected-electron
momentum distribution. For $M_L=1$, the total population in 2p emission channels beyond a distance
of 108 $a_0$ is 19.56\%,
whereas it is 1.06\% for 2s emission channels leaving C$^+$ in the 2s2p$^2$ $^4$P$^e$ state.
The total population in the outer region at the end of the calculation is 28.31\%, with 27.17\%
associated with the C$^+$ ground state and 1.07\% associated with the first excited state of
C$^+$. Higher-lying states of C$^+$ account for less than 0.25\% of the population in the outer
region, and therefore the contribution to the momentum distributions from outer-region channels
associated with these states is again assumed to be negligible. Again, the total population in
the outer region is noticeably larger than the population associated with ionization. It can be seen
that the total population in the outer region has increased by more than a factor 6 from the
population obtained in the $M_L=0$ calculation, with an increase in the emission
probability for a 2p electron by a factor 7. The emission probability for a 2s
electron has also increased, but only by a factor 2.

The reason for the difference in the ejected-electron momentum distribution is that emission
of $m_\ell=0$ electrons is strongly preferred in a linearly polarised laser field.
For an initial magnetic quantum number $M_L=0$, symmetry prevents the emission of an electron
with $m_\ell=0$, leaving C$^+$ in a 2s$^2$2p $^2$P$^o$ state with $M'_{L'}=M_L - m_\ell=0$.
Since transitions with $\Delta L=0$ are not allowed for $M_L=0$, the only 2s$^2$2p ionization
channels available are of the form 2s$^2$2p($^2$P$^o$)$\varepsilon\ell$ $^3$L with $\ell=L$.
However, the Clebsch-Gordan coefficient CG(1 L L; 0 0 0) is identical to 0. On the other hand,
for $M_L=1$, emission of a 2p electron with $m_\ell=0$ is allowed, and this process
dominates the ionization.

\begin{figure}%[tbh]
	\centering
		\includegraphics[width=9cm]{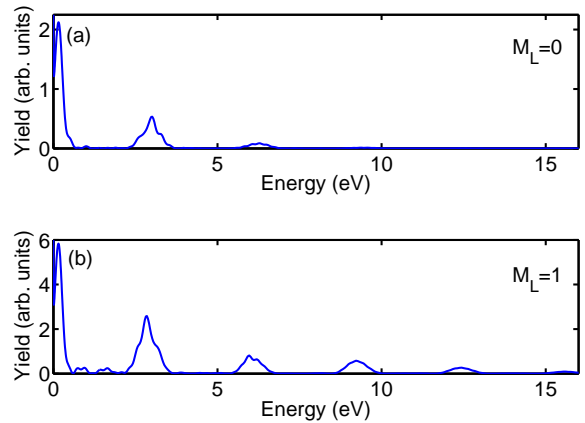} %{GSM0M1.png}%[width=300pt]
	\caption{(Color online) Photo-electron spectra for electrons leaving the residual C$^{+}$ ion in the
	ground state observed at an angle of 27$^{\circ}$ to the laser polarization axis for
	(a) $M_L$ = 0 and (b) $M_L$ = 1 in a laser field with a wavelength of 390 nm and peak
	intensity of 10$^{14}$ W/cm$^2$.}
          \label{fig:eground}
\end{figure}

The contribution of emission of a 2s electron to the total ionization demonstrates that
above-threshold ionization is important for C at this combination of wavelength and intensity.
It takes absorption of one extra photon to eject the 2s electron. This importance of
above-threshold ionization is also visible in the ejected-electron momentum distributions.
For emission of the 2p electron, three additional peaks at higher momentum can be seen for
$M_L=0$ and one for $M_L=1$. For emission of a 2s electron, however, another four peaks can
be seen along the laser polarization axis for $M_L=0$.

\begin{figure}%[h!]
	\centering
		\includegraphics[width=9cm]{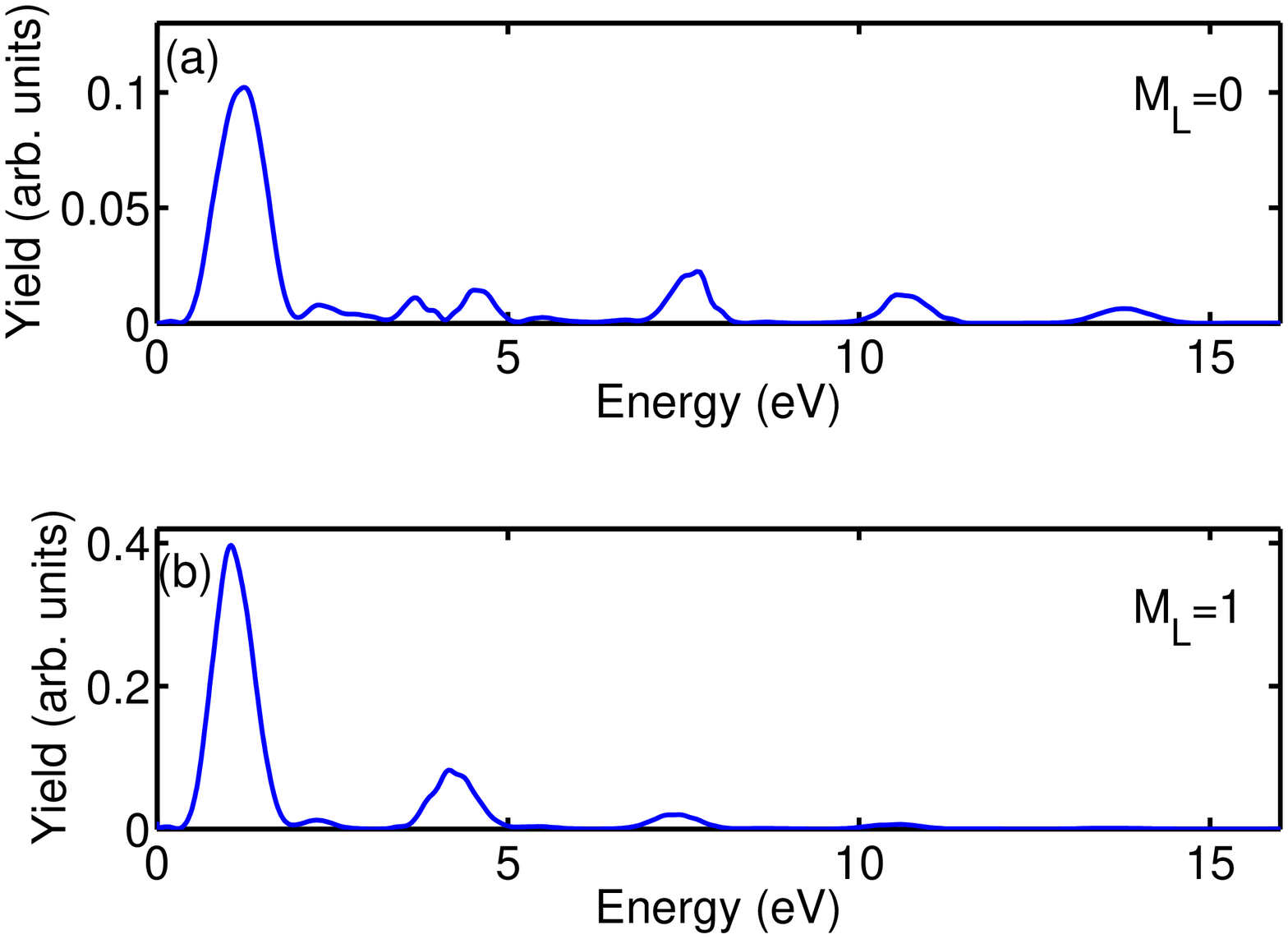} 
	\caption{(Color online) Photo-electron spectra for electrons leaving the residual C$^{+}$ ion in
	the excited 2s2p$^2$ $^4$P$^e$ state observed at an angle of 27$^{\circ}$ to the laser
	polarization axis for (a) $M_L$ = 0 and (b) $M_L$ = 1 in a laser field with a wavelength
	of 390 nm and peak intensity of 10$^{14}$ W/cm$^2$.}
\label{fig:eexcited}
 \end{figure}

To visualise the importance of above-threshold ionization in more detail, we can also investigate
the photo-electron spectrum for a particular emission angle.
The photo-electron energy spectrum associated with the 2s$^2$2p $^2$P$^o$ ground and $^4$P$^e$
excited state of the residual C$^{+}$ ion are shown in figures \ref{fig:eground} and
\ref{fig:eexcited}, respectively, for both $M_L$ = 0 and $M_L$ = 1. We have chosen an angle
of 27$^{\circ}$ with the laser polarization axis for these energy spectra, as this allows these
spectra to be studied with a reasonable magnitude for all 4 cases. 

Figure \ref{fig:eground} shows the photoelectron spectrum associated with the residual C$^+$
ion left in its ground state for initial C atoms in either $M_L=0$ or $M_L=1$. It can be
seen that the overall magnitude of the photoelectron spectrum is significantly larger for
$M_L=1$ than it is for $M_L=0$, which is primarily due to the increase in ionization yield
for $M_L=1$: the peak at 3 eV has increased by about a factor 5. It can furthermore
be seen that the number of above threshold peaks is substantially larger for $M_L=1$ than for
$M_L=0$. For $M_L=0$, the above threshold ionization peaks drop off rapidly
with increasing energy. For $M_L=1$, however, after a sharp drop for the second peak, the
drop-off for higher peaks is significantly slower. The origin probably lies with the
dominant $m_\ell$-value of the ejected electron. For $M_L=0$, this is $m_\ell=\pm1$, but it
is $m_\ell=0$ for $M_L=1$. Since $m_\ell=0$ electrons respond more strongly to the
laser field than $m_\ell=\pm1$ electrons, it is not unexpected to see additional absorption
peaks in the 2p emission spectrum for $M_L=1$.

Figure \ref{fig:eexcited} shows  the photoelectron spectrum associated with the residual C$^+$
ion left in its lowest excited state for initial C atoms in either $M_L=0$ or $M_L=1$.
This figure demonstrates that the probability of emission of a 2s electron has increased
for $M_L=1$ compared to $M_L=0$, even though the 2s emission is not visible
for $M_L=1$ in figure \ref{fig:elecmom1}. The structure of the ejected-electron energy spectrum
has changed, with the peaks for above-threshold ionization now
less visible for $M_L=1$ compared to $M_L=0$. In addition, the peak at 4 eV appears to be
affected by an interference effect for $M_L=0$, whereas no such effect is seen for $M_L=1$.
The origin of this interference structure is unclear. The inclusion of higher-lying thresholds
ensures that C resonances can be present at this energy, but only states of $^3$S$^e$ symmetry
would be accessible for $M_L=0$ and not for $M_L=1$.

The ejected-electron momentum distributions shown in figures \ref{fig:elecmom0} and
\ref{fig:elecmom1} can be used to
model the influence of spin-orbit dynamics. In experiment \cite{Hul13,Ekl13}, an electron
is detached from C$^-$ to leave a residual C atom. This C atom is in a superposition of the
different $J$-levels of the 2s$^2$2p$^2$ ground state, and this superposition will
evolve in time due to the spin-orbit splitting of the different $J$ levels. The evolution of
the C atom is measured experimentally by examining the dependence of the emission
of high-energy electrons along the laser polarization axis on time delay. The polarization axis
of the C$^-$ photodetachment laser is varied between perpendicular and parallel polarization,
and the dynamics is determined by examining the difference in C$^+$ yield along the probe laser
polarization axis between the two polarizations for the pump laser.

In our theoretical model, we assume that the pump laser detaches the $m_\ell=0$ electron from the
2p$^3$ $^4$S$^o$ ground state of a initial C$^-$ ion, leaving a C atom in its ground state with
$M_L=0$. For parallel polarization of the
probe laser, the initial state at zero time-delay is thus an $M_L=0$ state, whereas for
perpendicular polarization, the initial state is in a superposition of $M_L=1$ and $M_L=-1$.
We can then project the $^3$P$^e$ initial state with quantum numbers $L,M_L, S, M_S$ onto the
$J, M_J$ sublevels. We then propagate these $J,M_J$ sublevels in time according to the energy
splittings given in the literature \cite{Moo93}. After a time delay, we project back onto the
$L, M_L, S, M_S$ sublevels and use the population in the different $M_L$ levels to
predict the ejected-electron momentum
distribution. The emission within a cone of 11.7$^\circ$ from the laser polarization axis, with
a minimum magnitude for the momentum of 0.4 atomic units, similar to experiment \cite{Hul13,Ekl13}, is then taken as
the high-energy
ionization yield along the laser polarization direction
for both the initial state in $M_L=0$ or in the superposition for $M_L=\pm1$.  The propagation of the C ground
state during the time delay can give rise to three beat periods in the picosecond range,
$\tau_{1}$ = 1235 fs, $\tau_{2}$ = 769 fs, and $\tau_{3}$ = 2034 fs. 

\begin{figure}%[h!]
	\centering
		\includegraphics[width=9cm]{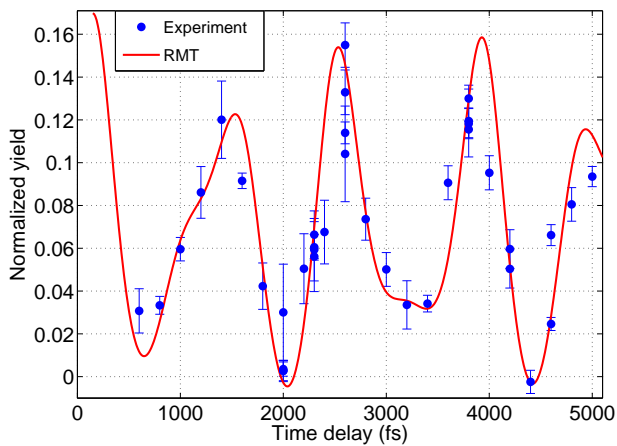}%[width=300pt]
	\caption{(Color online) Normalized high-energy ionization yield differences as a function of time delay for C.
	The experimental data (circles, \cite{Ekl13}) shows normalised ionization yield differences
	for energetic electrons along the probe-laser polarization axis between a probe pulse with
	 perpendicular polarization direction and parallel polarization direction. The theoretical
	 data shows ionization yield differences for energetic electrons along the probe-laser
	 polarization axis between C atoms with $M_L=1$ and $M_L=0$ at zero time delay.}
\label{fig:timedelay}
 \end{figure}

Figure \ref{fig:timedelay} shows re-scaled normalised high-energy ionization yield differences, similar to the
definition used in the experimental study \cite{Ekl13},
\begin{equation}
        \label{equation1}
			S(\tau)=\frac{S_{\pm1}(\tau)-S_{0}(\tau)}{S_{\pm1}(\tau)+S_{0}(\tau)},
\end{equation}
where $S_{\pm1}(\tau)$ is the high-energy ionization yield along the laser polarization axis obtained for the
initial superposition of $M_L=\pm1$ at a time delay $\tau$, and $S_0(\tau)$ the same %ionization yield
obtained for $M_L=0$. The high-energy ionization yields have been rescaled to match up with the experimental
results in overall variation of the normalised yields, zero yield position,
and zero time delay. The first reason for this rescaling is that the experimental study was carried
out at signficantly different experimental conditions: the wavelength was
1310 nm, and the peak intensity was estimated to be 4 $\times$ 10$^{14}$ W/cm$^2$. This may lead to
significant differences in the actual ionization yields. The second reason is the
assumption of pure $m_\ell=0$ emission by the pump pulse. Although the emission of $m_\ell=0$ electrons
should dominate, some $m_\ell=\pm1$ electron will also be emitted. However, these differences will
not affect the fundamental character of the dynamics
induced by the spin-orbit interaction: the same dynamics should be observed in both theory and
experiment.

%We note that the behaviour of $S(\tau)$ will no longer be
%described accurately by a superposition of trigonometric functions due to the division by
%$S_{\pm1}(\tau)+S_{0}(\tau)$, which varies with time delay.

Figure \ref{fig:timedelay} indeed shows excellent agreement between the experimentally observed
effect of spin-orbit dynamics on the ejected-electron momentum
distributions and the theoretical model. This demonstrates that it is appropriate to separate the
dynamics induced by the laser field from the dynamics induced by the
spin-orbit interaction for this particular investigation. The good agreement further demonstrates
that the main physical reason that the spin-orbit dynamics can be observed
in this scheme, is that the emission of 2p electron with $m_\ell=0$ is forbidden when the C atom
is in a state with $M_L=0$.

\section{Conclusions}
\label{sec:conclusions}

We have developed capability within the R-matrix with time dependence approach to investigate
multiphoton ionization of general atoms with non-zero initial magnetic quantum number. The size
of the calculations approximately double due to the need to take both parities into account for
each angular momentum. We demonstrate the capability of the
approach by investigating multiphoton ionization of ground-state C atoms at a wavelength of 390
nm. The ejected-electron momentum distributions show that both 2p and 2s electrons can be
ejected during the process with the emission of 2s electrons gaining in importance for initial
$M_L=0$ compared to $M_L=1$. For an initial state with $M_L=0$, emission of 2p electrons can not occur
along the laser polarization axis, and this provides an excellent means of demonstrating the
dynamics induced by spin-orbit coupling in the C ground state. A theoretical model of this
dynamics shows excellent agreement with the experimentally observed dynamics.

In the calculations, we have used significant expansion lengths in the description of C. We
have used 8 target states and a maximum angular momentum $L_{\rm max}=53$. The calculations
demonstrate that the RMT codes are capable of handling larger basis sets and more extensive
angular-momentum expansions, including outer-region expansions of
well over 1000 channels. Application of the RMT codes to problems involving significant CI
expansions in the inner region is therefore possible. This would be important for the detailed
treatment of ultra-fast processes in inner shells, where changes to the outer-electron orbitals
following inner-shell emission may have to be included.  

\section{acknowledgements}

 The
authors would like to thank M. Eklund and I.Yu. Kiyan for providing their data in numerical
form. They would further like to thank J.S. Parker and K.T. Taylor for
valuable discussions. The RMT codes were primarily developed by M.A. Lysaght and L.R. Moore.
This research has been supported by the EU Marie Curie Initial Training Network CORINF and by
the UK Engineering and Physical Sciences Research Council under grant no EP/G055416/1.This work
made use of the facilities of HECToR, the UK's national high-performance computing
service, which is provided by UoE HPCx Ltd at the University of Edinburgh, Cray Inc. and
NAG Ltd, and funded by the Office of Science and Technology through EPSRC's High End
Computing Programme.

\end{document}